\documentclass[12pt,a4paper,twoside,USenglish]{article}

\usepackage[margin=2.5cm]{geometry}
\usepackage[utf8]{inputenc}				

\usepackage{lmodern}
\usepackage{microtype}

\usepackage[numbers]{natbib}
\usepackage{hyperref}
\usepackage{url}
\usepackage{fancyhdr}
\usepackage{lipsum}
\usepackage[final]{changes}
\usepackage{booktabs}
\usepackage{multicol}
\usepackage{anyfontsize}

\newcommand\contributor[5]{%
  \goodbreak
  {\small
   \if!#4!\else\leftskip30mm\fi
   \noindent
   \if!#4!\else
     \ifx\f@rmat b\vrule\@width\z@\@depth29.8mm\fi
     \llap{%
       \if!#5!\else\expandafter\smash\fi
         {\lower29.8mm\hb@xt@26mm{%
           \includegraphics[width=26mm,height=32mm]{#4}\hss}\kern4mm}}%
   \fi
   \parbox[t]{\dimexpr\hsize-\leftskip}{%
     \raggedright
     \bfseries#1\par
     \mdseries#2%
     \if!#3!\else
       \ifx\f@rmat b\par\else,\space\fi
       \bfseries#3
     \fi}%
   \par\vskip\baselineskip
   \if!#4!\else
     \RaggedRight
     \ifx\f@rmat b\else\leftskip30mm\fi
   \fi
   \noindent#5\par}%
  \vskip2\baselineskip
  \def\@tempa{multicols}\ifx\@currenvir\@tempa\vskip\z@\@plus\baselineskip\@minus0.5\baselineskip\fi}

\usepackage{lipsum}

\begin{document}

\title{The European Commitment to Human-Centered Technology: The Integral Role of HCI in the EU AI Act's Success}

\author{André Calero Valdez, Moreen Heine, Thomas Franke, \\ Nicole Jochems, Hans-Christian Jetter, Tim Schrills\\ \\ University of Lübeck \\ Institute of Multimedia and Interactive Systems,\\23562 Lübeck \\ Germany}

\date{}

\maketitle
\abstract{
The evolution of AI is set to profoundly reshape the future.
The European Union, recognizing this impending prominence, has enacted the AI Act, 
regulating market access for AI-based systems.

A salient feature of the Act is to guard democratic and humanistic values by focusing regulation on transparency, explainability, and the human ability to understand and control AI systems. 
Hereby, the EU AI Act does not merely specify technological requirements for AI systems. The EU issues a democratic call for human-centered AI systems and, in turn, an interdisciplinary research agenda for human-centered innovation in AI development.

Without robust methods to assess AI systems and their effect on individuals and society, the EU AI Act may lead to repeating the mistakes of the General Data Protection Regulation of the EU and to rushed, chaotic, ad-hoc, and ambiguous implementation, causing more confusion than lending guidance. 

Moreover, determined research activities in Human-AI interaction will be pivotal for both regulatory compliance and the advancement of AI in a manner that is both ethical and effective. 
Such an approach will ensure that AI development aligns with human values and needs, fostering a technology landscape that is innovative, responsible, and an integral part of our society.}

\section{The increasing importance of AI}


In the rapidly evolving landscape of artificial intelligence (AI), the European Union's AI Act emerges as a pioneering legislative framework, aiming to safeguard human values and ensure the safe utilization of AI technologies. This legislative initiative categorizes AI systems into four distinct risk levels, with each category subject to specific compliance criteria. However, the operationalization and implementation of these criteria present significant challenges, underscoring the need for a meticulous examination of the Act's provisions and their practical implications.

The objective of the present paper is to critically analyze the EU's AI Act, focusing on the ambiguities and challenges inherent in operationalizing and assessing the compliance criteria. It delves into the intricacies of the certification processes, shedding light on the uncertainties surrounding the authorities responsible for verification and the methodologies that need to be employed for assessment. The EU AI Act's interpretative challenges, particularly concerning provisions related to human oversight, safety, and transparency, are scrutinized, highlighting the subjective (i.e., variably interpretable) nature of these criteria and the potential inconsistencies in their application and enforcement.

Furthermore, our present paper underscores the pivotal role of Human-Computer Interaction (HCI) in the context of the AI Act. It posits that the requirements for human oversight, safety, and transparency are intrinsically linked to human perception and interpretation of AI, thereby placing HCI at the forefront of AI development. We argue for the integration of HCI principles in the development of AI systems, emphasizing that fulfilling the Act's criteria necessitates a comprehensive understanding of human limitations and an adherence to fundamental HCI values.

\begin{figure}
    \centering
    \includegraphics[width=1\linewidth]{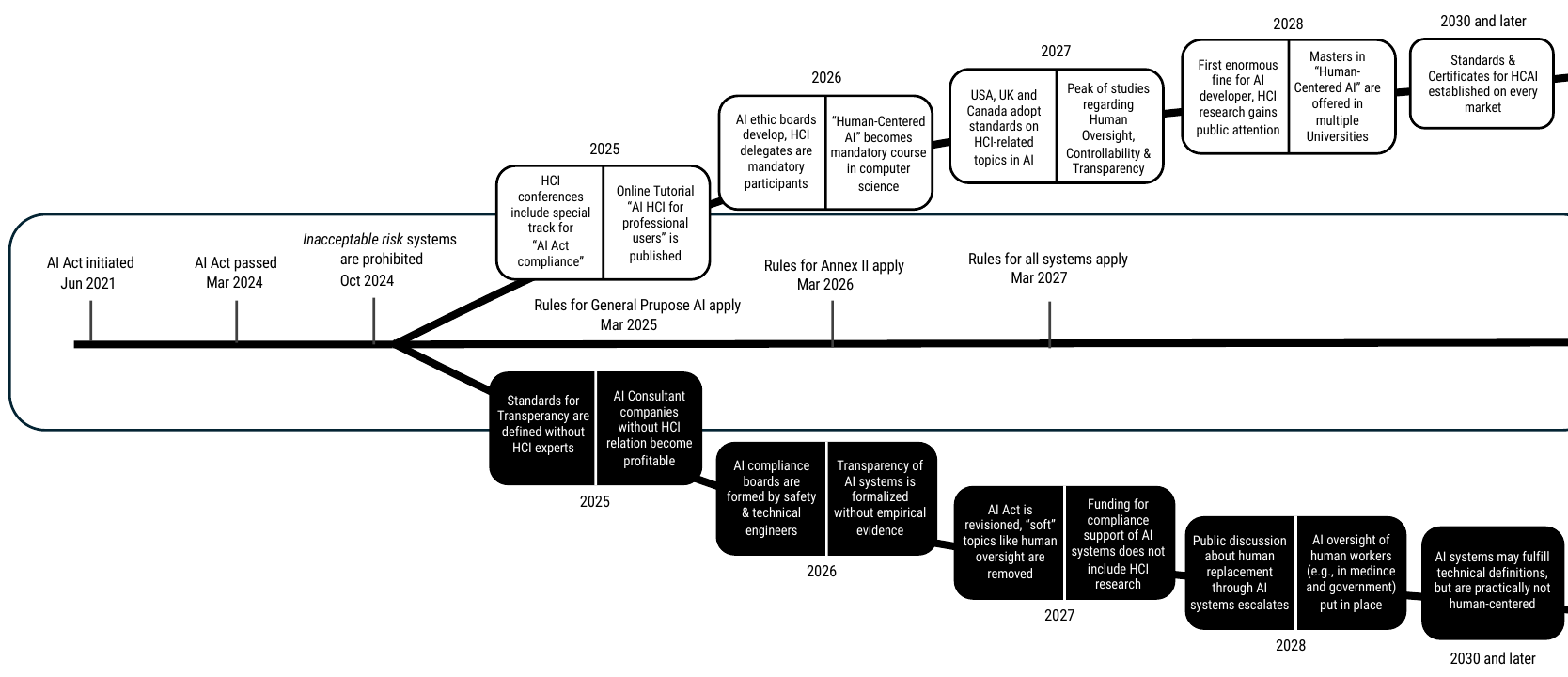}
    \caption{Possible timelines with potential events either addressing or ignoring the importance of HCI in relation to the AI Act.}
    \label{fig:timeline}
\end{figure}

In summary, our present paper provides an examination of the EU's AI Act. First, we highlight the challenges in operationalizing its criteria. Second, we discuss the ambiguities in certification and verification processes. Finally, we describe why HCI plays a crucial role in shaping the future of AI development. Through our analysis, we seek to contribute to the ongoing discourse on responsible AI and emphasize the imperative of aligning AI technologies with human-centric values and societal norms. 
The field of HCI must start to address these challenges in the next five years to remain in the loop in decision-making about the use of AI.
\added{
We conjecture two different timelines as possible outcomes where the field of HCI either embraces its role in concretizing the AI Act or continues with business as usual failing to live up to the implicit call to action in the AI Act (see Fig. \ref{fig:timeline}).
}

\section{The EU AI Act}

The European regulation of Artificial Intelligence (AI) aims to strengthen research and industry while simultaneously ensuring safety and fundamental rights by focusing on excellence and trust. The legal text is currently still in the final stages of the legislative process. However, the resulting activities in the standardization process are not yet finalized. The formulation of the regulations is deliberately abstract. This is intended to ensure flexibility, broad applicability, adaptability to changes, and the avoidance of legal loopholes. 

The AI act relies on risk-based regulations, which have to be specified through codes of practice. These codes are developed in cooperation between industry, academia, civil society, and the commission, supported by independent scientific expert panels responsible for classifying and reviewing AI models and issuing risk warnings. A new AI Office within the European Commission will ensure the development of AI policy at the European level and monitor the execution of the forthcoming AI Act. To promote innovation and test practical applications, AI regulation also includes experimental clauses and the establishment of regulatory sandboxes. These allow AI systems to be tested under real conditions in controlled environments without neglecting regulatory standards.

\subsection{Risk classes in the AI act}

The legislative framework classifies safety-critical areas into four distinct categories, each with its own set of regulatory requirements.

The AI Act presents a structured framework for regulating AI applications by categorizing them into different risk levels, namely 1) minimal, 2) high, 3) unacceptable, and 4) specific transparency risks. This categorization informs the regulatory requirements necessary to achieve compliance \added{(see Table~\ref{tab:ai_risk_levels})}.

\begin{table}[h!]
\tiny
\centering
\begin{tabular}{p{1.8cm}p{3.0cm}p{3.2cm}p{2.9cm}p{2.9cm}}
\toprule
\textbf{} & \textbf{Unacceptable Risk} & \textbf{High Risk} & \textbf{Limited (Transparency) Risk} & \textbf{Minimal Risk} \\ \midrule
\textbf{Criteria} & AI systems that threaten safety, livelihoods, and rights of individuals & AI systems used in critical areas, where significant risk to health, safety, or fundamental rights exists & AI applications with limited risk to rights or safety & AI applications with minimal or no risk\\ \midrule
\textbf{Market Entry Barriers} & Prohibited from the market. No entry allowed. & Mandatory compliance checks, rigorous testing and certification, data governance and accuracy standards, detailed documentation, human oversight, and periodic audits. & Obligation to inform users when interacting with an AI, transparency in data usage, easy opt-out mechanisms. & No additional barriers beyond existing laws. \\ \midrule
\textbf{Specific \mbox{Examples}} & Social credit systems, emotion recognition, AI exploiting vulnerabilities, behavioral manipulation, untargeted facial image collection, selective predictive policing, real-time biometric identification in law enforcement & Medical devices, vehicular technologies, HR management systems, educational tools, electoral influence tools, critical infrastructure management, law enforcement tools & Requires transparency but generally includes applications that are interactive without significant consequences; e.g., chatbots or image editing & Generally includes all applications with negligible risk to users and do not require specific regulatory oversight; e.g., video games or spam filters  \\ \midrule
\textbf{Regulatory Requirements} & Complete prohibition & Fundamental rights and conformity assessments, registration in public EU database, risk and quality management systems, strict data governance, mandatory transparency, human oversight, high standards of accuracy, robustness, cybersecurity, consistent testing and monitoring & Transparency in data usage, user interaction notifications & Compliance with general consumer protection standards, GDPR compliance \\ \bottomrule
\end{tabular}
\caption{Summary of AI risk levels, market entry barriers, and regulatory requirements under the EU AI Act with specific examples}
\label{tab:ai_risk_levels}
\end{table}

AI applications classified under the unacceptable risk category encapsulate technologies perceived as antithetical to societal values and individual liberties. This includes, but is not limited to, systems like social credit scoring, emotion recognition in workplaces and educational settings, AI designed to exploit vulnerabilities such as age or disabilities, and technologies capable of behavioral manipulation or subversion of free will. Moreover, this category encompasses the untargeted collection of facial imagery in public domains, public space facial recognition, biometric categorization mechanisms, selective predictive policing (with certain exclusions), and the tightly constrained use of real-time biometric identification in law enforcement contexts.

\added{Alternatively}, the high-risk category is reserved for AI systems with a substantial propensity to inflict harm on human safety, health, environmental integrity, or property. This broad category encompasses a diverse array of applications ranging from medical devices, vehicular technologies, human resource management systems, educational tools, mechanisms influencing electoral behaviors, to the management of critical infrastructures, and tools employed in law enforcement sectors.

Pertaining to high-risk AI, the regulations stipulate a series of stringent compliance measures. These encompass conducting fundamental rights and conformity assessments, mandatory registration in a designated public EU database, and the institution of both risk and quality management systems. 
A paramount requirement for these AI systems is adherence to strict data governance protocols aimed at bias mitigation and ensuring data representativeness. Transparency is a critical mandate, necessitating the provision of lucid documentation and instructional materials. Furthermore, human oversight is a crucial requirement, mandating the availability of auditable logs and, potentially, system explainability. These systems are also obligated to meet elevated standards of accuracy, robustness, and cybersecurity, 
underscored by consistent testing and monitoring practices.

Systems that pose little or no risk to citizens' rights or security fall into the minimal risk category, such as AI-enabled video games or spam filters. \added{When consequences of AI use are predictable, reversible, and inherently pose little risk, no regulatory requirements are imposed.
Such systems still have to comply with general consumer protection standards and the GDP, but no further restrictions apply to them with regard to the AI Act.}

\subsubsection{General purpose AI}
In an exceptional category, General Purpose AI, especially those employing foundation models, is distinguished due to its inherent risk factors, necessitating augmented transparency and comprehensive disclosure protocols.

For foundation models, characterized by a significant investment in computational effort (exceeding $10^{25}$ FLOPS), there are additional transparency requirements encompassing technical documentation, training data oversight, and protective measures for intellectual property rights. Foundation models that carry systemic risks, exemplified by platforms such as ChatGPT, are subject to more stringent measures, including model evaluations, supplementary risk assessments, adversarial testing, and the establishment of incident reporting mechanisms. All content and interactions generated by generative AI must be explicitly labeled and be discernible to human users.

\subsection{Enforcement and compliance}

To ensure regulatory compliance, the AI Act imposes substantial penalties for breaches. Entities may incur fines up to 7\% of their global turnover or €35 million for the deployment of AI in unacceptable applications and up to 3\% of global turnover or €15 million for other infractions, with tailored provisions for SMEs and startups.

The enforcement framework is comprised of an EU AI office, an AI board, provisions for individual complaints, and the establishment of market surveillance authorities within member states. This comprehensive apparatus is designed to monitor and regulate AI applications across the entire spectrum of assessed risk levels, ensuring a harmonized and effective governance of AI technologies within the jurisdiction.

\added{
There are some parallels with the EU General Data Protection Regulation. Both laws have global ambitions and entail serious sanctions. The question of to what extent the AI Act can promote or inhibit innovation, especially for SMEs and organizations working for the common good, arises again. This issue is closely linked to a clear mandate for scientific research that is societally relevant and applicable (similar to the GDPR with regard to IT security research): How can the socio-technical systems in question be designed to meet the ambitious development and operational conditions in a resource-efficient manner, even amidst technological advancements?
}

\subsection{Transparency and human oversight in regulations}
A particular focus of the AI Act is on transparency and human oversight. 
These are important criteria as they tie the regulation to those most affected by any regulation---humans.
However, the terminology applied \added{in the AI Act} is (partially) unexpectedly deviation from current research in explainable AI (XAI). 
Terms such as \textit{explainability}, \textit{interpretability} or \textit{understandability} have been a key research interest in recent years. However, they inadvertently tie the underlying quality (explainability) to human (subjective) evaluation, as in ``Does the application explain the decision to me?'' or ``Do I understand the decision?".
Transparency, on the other hand, is a term associated with physical properties of objects and alludes a more objective interpretation of the desired quality of AI systems. This allows a technical interpretation of transparency, as in,
``Can we find/define a measurement tool that demonstrates transparency?''
Human oversight or controllability---as used in the regulation---is much closer to human evaluation and ties into existing well-established research areas in HCI and human factors research.
As both these aspects are key aspects of the regulation we must assume that the goals of the AI Act are intended to be not only a mere technical interpretation of the aforementioned goal criteria. Nevertheless, these still underdefined concepts beg the question: How can we attain these criteria?

\section{There is no reliable AI regulation without a sound theory of human-AI interaction}

A major concern with the AI Act is that many terms and concepts are not sufficiently detailed, since sound theoretical models to describe human-AI interaction are still in development. For example, the AI Act requires high-risk systems to provide information about an AI system's accuracy and limitations, but it does not specify how detailed this information must be or what methods can be used. With fines roughly twice as high as the GDPR, developers of AI systems are demanding less ambiguity on their obligations to comply with the law. 

But what is needed for the AI Act to be an impetus for innovation rather than ambiguity? Following Lewin's maxim that ``there is nothing as practical as a good theory''~\cite{lewin1943psychology}, a deep theoretical understanding of human interaction with AI systems is needed to realize the potential of regulation as a driver of innovation. Thus, to support a successful environment for AI development, the HCI community must incorporate existing theoretical concepts from psychological science, especially on human action regulation and experience in technology-rich/artificial environments (i.e., engineering psychology), into education of developers and users, research, and product development. 

\subsection{Humans do more than just using AI systems}

In its most recent version, the EU AI Act aims to ``promote the uptake of human-centric and trustworthy artificial intelligence''~\cite{AIACT}. It also requires from high-risk systems to enable human oversight~\cite{AIACT}, by supporting users in understanding the AI system, staying aware of potential biases, and allowing them to correctly interpret AI systems' output. The EU AI Act thus also calls on us as the scientific community (and general public) to view people not just as the executing instance who operates a technical system, but as an independent, responsible entity with requirements to responsibly handle the information processing tasks at hand. 

First, when human users aim to interact with an AI system, they assess how the system processes information, i.e., develop an information processing awareness~\cite{schrills2023users}. While we do not want to discuss the notion of (situation) awareness in detail here, \added{it is important} to highlight, that it is a construct about a state of the human user~\cite{endsley1995toward}. In comparison to existing constructs in research on AI systems, for example, the Explanation Satisfaction Scale (ESS,~\cite{hoffman2023measures}), the concept of situation awareness addresses how AI interaction changes human perception and behavior. The ESS aims to evaluate system properties, but is, in the end, focused on the system and not its user. However, we postulate that users regulate their actions---for example, when making a decision---based on the before-mentioned information processing awareness. That is, the extent to which users perceive, understand, and predict an AI system's information processing~\cite{schrills2023users} is the basis for their ability to oversee this system. 

Second, many AI systems are and probably also will be developed for tasks or processes, that have been carried out by human operators before (which is, after all, the definition of automation~\cite{onnasch2014human}). On top of that, depending on the field they are developed for, their integration into human tasks will potentially be limited~\cite{AIACT}. Therefore, a fruitful approach could be to take human action control as a starting point and view the gradual integration of automated information processing as an adaptation of this action control. This also allows us to address crucial scientific and ethical questions: at what point is control over the action so far removed from humans that they can no longer or should no longer be able to take responsibility for it? At what interaction points in a shared task can human oversight be established by improving human awareness of information processing?  

Third, in light of the AI Act, it is immensely important to note, that human users of AI systems may approach these systems with biases and a limited ability \added{to process all presented information cognitively}. That is, the perception of risk may be different when users have a sufficient feeling of control compared to situations where they \added{do} not feel in control~\cite{sjoberg2004explaining}. In addition to this, huge amounts of information may convince users to positively assess outcomes even without verifying their accuracy~\cite{ferguson2022explanations}. When humans are modeled as rational, perfect users of AI systems, such biases and ergonomic requirements will not be mitigated through AI systems but potentially enhanced---or, in the worst case, abused. 

All in all, the EU AI Act is a call to focus on a deep understanding of human users of AI---including potential human biases, emotional, motivational, and cognitive states of human users, and, ultimately, the conditions under which human oversight and responsibility are possible---or not. 

\subsection{AI Systems constitute automated information processing}

Regarding the definition of AI systems in the AI Act and in the OECD definition~\cite{Yeung2020-nq}, an important aspect of these systems is that they process information autonomously, and their results can be decisive for human action regulation. This characterizes AI systems as types of automation of information processing that are characterized by autonomy and connection to human actions. Hence, to discuss AI systems' effect on human users, they can be anchored in theoretical concepts of automation.

\added{Key considerations include the suitability of automation for various tasks and its impact on human activities. It is essential to understand how the ergonomic design of AI systems differs from less automated systems in order to develop AI that enhances user capability. Accurate models of human response to AI are crucial for creating 'trustworthy AI' that aligns users' expectations with reality. Achieving this balance is vital to prevent over-reliance, also known as complacency, in automation}~\cite{wickens2015complacency}. 
However, the integration of AI systems into human information processing can fail when their utilization is hampered by (unwarranted) mistrust~\cite{brauner2019happens}. We can assume that AI systems are not only experienced differently depending on their design, but may actively change how human users approach a task~\cite{vanDongen2013}. That is, the integration of AI into human work modifies how users regulate their actions and which information they use to make decisions, requiring careful consideration of how humans control and interact with automated systems. 

The rich literature on signal detection and alarm systems, for example, can provide insights for AI design, particularly in ensuring effective communication of critical information to enhance safety and performance~\cite{onnasch2014operators}. 
Viewing AI through the lens of automation allows for the use of empirical research to develop technologies that can augment human capabilities. On top of that, \added{this perspective} allows predictions about the AI systems' influence on human performance, accountability, and the human ability to exert control~\cite{rottger2009impact}. 
\added{These considerations are critical as human control and oversight are key regulatory requirements for higher-risk classes of AI in the AI Act.}

\subsection{AI Systems can have different levels of automation}

To characterize diversity in automated systems, the concept of ``levels of automation'' has been well-established and extensively discussed in human factors and engineering psychology for decades. This perspective allows for the classification of different levels of autonomy in systems, explicitly balancing the level of control retained by human users against the autonomous actions of the system. 
For example, as one of the most-cited approaches, Parasuraman et al., formulate ten different levels of automation, ranging from manual operation to full autonomy. Their work provides a nuanced approach to understanding the extent of function allocation and to some degree to the structure of interaction between humans and automated systems~\cite{parasuraman2000model}.

Conceptualizing AI systems in ``levels of automation'' frameworks may enable developers and regulators to dissect the complexity of AI systems, such as distinguishing between recommender systems and decision support systems, which may have different levels of automation and consequently different implications for user decision-making and information processing (see~\cite{tatasciore2022should}). 
That is, levels of automation can be a framework that supports developers to understand how AI systems can be improved. Contrary to a risk classification, levels of automation can be associated with empirically tested solutions instead of legally required documentation. 
It is important to note, that the level of automation should not be as high as possible, but fit the context of an AI's deployment and be designed to enable optimal joint human-AI performance. For example, Miller argues that recommender systems (possibly representing higher levels of automation) may hamper the user's ability to make decisions in comparison to evaluative AI, which provides information~\cite{miller2023explainable}. Understanding these distinctions is important to address accountability and control within AI systems, aspects that are closely aligned with the objectives of the EU AI Act to promote transparency, security, and users' rights in the use of AI technologies~\cite{AIACT}.

Hence, in research, education, and professional practice, characterizing the specific level of automation of an AI system becomes a fundamental question. Questions such as ``what level of automation does my AI system possess?'' and ``what level of control and autonomy is granted to the user?'' are essential to ensuring that AI systems are designed, implemented, and governed in an ethical, user-centered, and compliant manner. Thus, it is a crucial task for HCI research to revisit research on human-automation interaction and apply it to the design, study, and development of AI systems. In this process, this will fill the concepts of the AI Act with life.

\subsection{Research is only as reliable as its ability to develop theory}

In the face of the EU AI Act, HCI research does not necessarily need to create new theories, tools, or methods, as previous research on human reaction to automation is highly transferable. Problems identified in automation research, such as \added{opacity/lack of transparency}, illustrate how existing frameworks can guide the development of more transparent and user-friendly AI. However, AI introduces new challenges that require the application of past research for effective solutions. For example, while the concept of Situation Awareness has been used in research on automated systems, more specialized concepts like Information Processing Awareness can prove useful to explicate and examine the challenges of AI systems~\cite{schrills2023users}. 

All in all, the key challenge for HCI research was, is, and remains to develop theories on how users construct mental models of AI systems~\cite{johnson2010mental}, and how feedback and learning influence such models. Research on AI has demonstrated, that users might be convinced of a system without understanding it~\cite{chromik2021think}. Existing research on explainability pitfalls or even dark patterns in explainability~\cite{ehsan2021explainability} demonstrate: Systems containing more complex information processing can be a barrier for efficient feedback and learning of users in comparison to existing, automated systems and explanations could even be weaponized to convince users of an AI system's accuracy. In addition, improving how diagnostic information is selected and presented to users may be more complicated in data-driven models~\cite{adadi2018peeking}, demonstrated by a rising number of causal research in AI systems~\cite{holzinger2021towards}.

In addition, refining collaborative actions between humans and AI based on interdependence and shared experience is critical to advancing human-centered AI. With a significantly increasing number of input parameters, interaction points, and feedback channels (e.g., through anthropomorphic features), the user and system exert influence over each other while cooperatively engaging in a task increases as well. The increasing coupling of humans and systems during the processing of information must be modeled more precisely in theory.

\section{There is no trustworthy AI without HCI}

The requirements for AI systems set out by the EU AI Act reflect an important perspective of the safety of AI systems: a human-centered view, that calls for empirical research on humans' reaction to AI systems. It is opposite to a purely technical view, defining concepts such as the trustworthiness or robustness of AI systems without integrating the human factor. The AI Act, striving to ensure the safety of AI systems in terms of their controllability and transparency by governmental, organizational, or natural individuals, raises questions about existing methods in AI systems, e.g.: how do explanations~\cite{adadi2018peeking} affect an individual's ability to detect errors? Which training can prevent complacency of novices or experts, and which information do users need to provide oversight? 

In 2023, the EU issued the European Committee for Standardization (CEN, Comité Européen de Normalisation) to provide standards for trustworthy AI. This standardization process is a crucial element of the EU's approach on operationalizing the EU AI Act. The EU's institutions took responsibility for defining terms, that are still highly debated in HCI research: trustworthiness, controllability, or human oversight are just a few examples. It is promising, for example, that the detectability of a system's state and a user's influence on changing the state, are central to defining what ``controllability'' of systems is (see for example~\cite{norman1986cognitive}). However, it is the HCI community, that needs to provide tools, methods, and---most important---replicable and theory-driven research to support the development of human-centered AI.

Historically, HCI research has successfully mastered comparable challenges, e.g., when faced with disruptive technologies such as Personal Computers and the World Wide Web. By understanding and formalizing formerly hard-to-grasp and fuzzy concepts such as ``user friendliness'', ``usability'', or later ``user experience'', HCI has not only helped to establish industry standards for testing and improving complex human-centered qualities of systems but also has arrived at a core insight about their nature: Unlike the technological parameters of a system such as storage size, response times, or energy consumption, human-centered qualities do not reside in a system but emerge from how specified users are using a system for achieving their goals in their specified context of use~\cite{cockton2004value,cockton2006designing}. Without understanding and observing real users in real-world contexts, the design and the assessment of the human-centered qualities of a system are bound to fail. In light of the coming AI revolution, it is now even more important that HCI faces that challenge with the same openness for---and insistence on---involving real-world users and considering their many different practices and contexts of use for achieving truly human-centered technologies.

In the next five years, students, researchers, and practitioners of HCI need to familiarize themselves with a new regulatory framework and contribute to its concretization. Here \added{and in the following sections}, we identify three core challenges for HCI research in the coming years: (1) developing metrics to evaluate human-AI interaction and its compliance with EU legislation, thereby enabling a continuous improvement of the AI Act, standards and regulatory monitoring; (2) conducting theory-driven research on instruction for users of AI systems and education fostering human abilities to control AI appropriately; and (3), the (continuous) development of multimodal designs and interaction patterns that enable developers to comply with the AI Act and support ethical values of human-centered technology. 

\subsection{HCI must define and disseminate methods to evaluate human-centered AI}

The scientific discourse around Human-Centered AI demonstrates how complex it is to define and operationalize related concepts, such as trustworthiness or control (see~\cite{speith2022review}). 
While EU legislation will not end the scientific discourse, it highlights the importance of providing reliable theories and methods to designers, developers, and deployers of AI systems. That is, two years after the AI Act comes into effect, deployers of AI systems (including governmental, educational, and industrial actors) need to be able to prove compliance. That is, a reliable and testable definition of AI compliant with the AI Act needs to be established, and with it, adequate methods to assess AI systems' trustworthiness. 

Previous, empirical research on trust has demonstrated that high levels of accuracy of AI systems alone are not sufficient to provide trustworthy AI systems~\cite{bansal2019beyond}. In order to develop reliable, trustworthy AI systems, developers must examine how users experience AI systems and how human-AI interaction may lead to complacency or unwarranted caution~\cite{shneiderman2020human}. Merely requesting explanations to be presented by AI systems or explainable AI (XAI) will not suffice. Undesired effects of explanations by XAI-system can happen, as has been demonstrated before~\cite{eiband2019impact, schrills2023users}. Unsurprisingly, the way XAI can improve human usage of AI systems is an emerging research topic~\cite{miller2023explainable}, which is however still in its infancy. 
The effect of explanations should therefore be well understood and---as stated in the AI Act---used in appropriate situations. Conversely, explanations should be applied with caution if they are qualified to limit human control (e.g., through information overload~\cite{ferguson2022explanations}). 

But what can HCI research do, given that requirements of trustworthy AI may be highly dependent on task, context, or even individual user characteristics? In the next years, the role of HCI will shift because core values of human-centered design---namely controllability and human oversight---are central to the AI act. Human-centered design has evolved from a requirement to ensure the safety of automated systems to the requirement of European Legislation itself. Accordingly, HCI must provide methods that are not only effective but reliable, transparent, and able to serve in certification processes (see also~\cite{shneiderman2020human}). In summary, HCI methods will move beyond research and development as they will become integral tools of EU legislation.

\subsection{HCI must explore and quantify the impact of education and instruction on users of AI}

In addition to comprehensive documentation, the appropriate presentation of information---in the form of a manual---will be key to meeting the requirements of the AI Act.
However, recent HCI research of AI systems, especially XAI systems, demonstrated a focus on visual or textual information, which accompanies specific decisions (local methods, e.g., ~\cite{ribeiro2016should}), explains the model in general (global methods, e.g.,~\cite{sundararajan2020many}), or aims to integrate both approaches (glocal methods, e.g.,~\cite{achtibat2023attribution}). 
Considering limited human capabilities to process information and to understand algorithms, HCI research needs to examine how to train and prepare users to be sufficiently qualified to exert control over AI systems and take responsibility. That is, controllability is not only dependent on the transparency and design of an AI system but also on an individual's ability to utilize the given information. 

HCI research has two paths to support users' ability to exert control over AI systems and take appropriate responsibility for results: improving methods for education and examining how to instruct AI users. For the improvement of education in AI systems, existing approaches like simulations~\cite{schrills2023safe} and scenario-based design~\cite{sun2022investigating} must be tailored to challenges accompanying AI systems, e.g., high volumes of data and continuous learning and adoption. Furthermore, learnings from the field of automation need to be adopted, e.g., how mental models of users can be improved to avoid breakdowns~\cite{benner2021you} such as the first failure effect~\cite{parasuraman1993performance}. But how can HCI research---and respectively EU authorities---define the requirements for users before they can (legally) operate high-risk AI systems? As this question will shape the future of work, HCI research needs to design valid educational approaches, addressing users with diverse characteristics~\cite{franke2019personal}.

\subsection{HCI must develop solutions for human-centered AI}

Data-driven methods of Machine Learning present developers with more challenges when it comes to transparency, traceability, and controllability (see~\cite{adadi2018peeking}). Accordingly, the HCI community has engaged in the development of techniques to provide these features, even in systems utilizing deep learning or Large Language Models. For example, the visualization of weights of input parameters via Shapley Values~\cite{sundararajan2020many} can support users in identifying (undesired) biases in data-driven models. The visualization of Shapley values, however, can be done in various ways (see~\cite{cooper2021explaining}). Which visualization is the correct one, given a human's task to identify non-compliant behavior? Which visualization may convince human users of a system's fairness, but does not support them in detecting unfair decisions? Adequate presentations of machine learning models for a huge diversity of users will require creative and human-centered design processes in the coming years. 

On top of that, while many approaches to support human-centered AI are visual or textual, the HCI community needs to put its experience of interactions to work. When aiming for high levels of user controllability, enriching AI interaction with additional information is not enough. For example, the control of generative AI systems can be highly customized through a variety of slider-based options~\cite{dang2022ganslider}, but potentially overburdens users. Especially in the context of general purpose models, designing how users can control their functionality, will be a central research topic. The ergonomics of the systems will depend on how well the complexity of the underlying models can be abstracted and users can be given control over them in (sequential) interaction. Existing studies show that even small changes in the sequence of interaction can lead to changes in human behavior~\cite{vanDongen2013}.

\subsection{Two visions for HCI}
\added{
Possessing a robust theoretical foundation in Human-AI interaction, coupled with the capability to objectively evaluate key facets of this interaction---such as transparency, controllability, and oversight---will constitute the essential critical thinking tools of the future (see Figure~\ref{fig:timeline}).} 

\added{
In an optimistic scenario, the field of HCI will eagerly adopt these theoretical frameworks and methodologies, weaving them seamlessly into both research and education pertaining to human-centric artificial intelligence curricula over the ensuing decade. Departments dedicated to HCI will allocate enhanced scrutiny to the regulatory dimensions of AI, while faculties of engineering and computer science will forge a closer amalgamation of HCI principles within AI research and educational efforts. With the activation of regulations delineated in Annex II of the AI Act, both the academic sphere and the industrial sector will be primed to affirm regulatory compliance and certify adherence to established norms. These regulations will align with public expectations, fostering a more profound understanding of Human-AI interactions, culminating in the refinement of AI system designs, the establishment of beneficial standards and certifications, and fostering a more collaborative interface with artificial intelligence systems.
AI systems, once operational, will enhance human cognitive capacities. Simultaneously, those who attempt to bypass regulations and contravene European values will face prosecution and punitive fines, as prescribed by the regulation.}

\added{In a less favorable vision, the field of HCI may find itself sidelined, overshadowed by a more technocratic implementation of the AI Act. Under such a scenario, the nuanced understanding of human factors in AI integration would be neglected in favor of stringent, technically-focused regulations that prioritize compliance over contextual relevance. This shift could lead to AI developments that, while regulatory compliant, are less attuned to the user-centric principles that enrich technology with humanity. Consequently, this could result in AI systems that are technically adept but deficient in fostering meaningful, ethical, and intuitive interactions with human users. 
In the bleakest scenarios, AI could potentially dominate human decision-making, especially in critical domains like medicine, despite technically remaining under human control. This could lead to situations where AI systems, driven by their programming and efficiency metrics, override human judgments and ethical considerations, prioritizing algorithmic outputs over human expertise and empathy. Such a development would not only erode trust in AI applications but also endanger fundamental human values by marginalizing the human element in vital decision-making processes.}

\section{There is no community without common language and communication}

The last large-scale regulation by the EU---the GDRP---caused tremendous changes in worldwide software applications. While the intention of the GDPR was to preserve privacy and allow control over data flows for end-users, most notably, the introduction of cookie banners on websites was the most visible effect of the regulation. Additionally, this change has furthermore induced other unintended ill consequences, such as the use of dark design patterns in Cookie consent forms~\cite{krisam2021dark}. 
Privacy disclaimers must be the most frequently presented and also most frequently ignored legal speech in existence, causing an economic burden of 2.3 billion USD annually~\cite{castro2014economic}.

While cookie banners and their atrocious usability were obviously not the intended outcome of the GDPR, unintended consequences of large-scale regulations---such as the AI Act---loom around the corner. Weaponized AI explanations, pro forma certification, or shifting the burden of liability to end users, are just a few examples of what could go wrong 
and threaten the success of the AI Act. 
So, how can we prevent a GDPR disaster 2.0? Who is capable of preventing similar unintended malicious consequences, and what tools will they need?

For the AI Act to become successful, the intended outcomes and the actual outcomes must align. And as shown above, the intended outcome of the regulation is to safeguard human values and ensure the safe utilization of AI. 
Measuring and ascertaining these outcomes cannot be conducted by engineering sciences and HCI alone. 
Other disciplines in SHAPE\footnote{SHAPE - Social sciences, Humanities, and the Arts for People and the Economy} and STEM\footnote{STEM - Science, Technology, Engineering, Mathematics}  will provide insight into \textit{what} to safeguard and into \textit{how} to safeguard these values.
Thus, it is critical to include other disciplines in this discourse and to provide a means of understanding what this discourse is about. However, discourse about AI can only be non-superficial and meaningful when the language used adequately represents the technological intricacies of the topic. Thus, a transdisciplinary language of AI is needed.

From an interdisciplinary perspective, different fields utilize different paradigms, methods, and theories with partial overlap. The fish-scale model of interdisciplinarity by Campbell~\cite{campbell2017ethnocentrism} posits that each discipline covers a certain aspect of our reality and that by overlapping different disciplines a full picture of reality may emerge. In this metaphor, HCI is the scale that overlaps with both the scale of engineering and the scale of social sciences and humanities. 
Thus, the responsibility of explanation is on the field of HCI. We as a community must demonstrate that our insights are able to translate between technological applications and value-oriented human interaction. We must translate and connect the quantitative engineering approaches to making AI transparent (e.g., Shapley values, LIME, or SHAP) with the qualitative meaning-seeking processes of the social sciences and humanities, that try to understand what transparency is for a heterogeneous group of individuals and communities. HCI is uniquely suited to be both a translator and an ambassador for human-centered AI to help both the engineering sciences create more transparent AI, and the humanities and social sciences to measure and reflect on the consequences of wide-scale AI utilization. 
To fulfill this role, we must pick our obligations to demonstrate that we will be able to do these roles justice. These obligations could be concrete research goals aligned with demands from the AI Act.
We must pick the most pressing research questions, even if we will only know how to answer them in five to ten years from now, we already do know that we should answer them now.
For example, should we as a field be able to demonstrate that the mandate for human control---as issued by the AI Act---is fulfilled by a certain user interface to AI-based systems? Otherwise, considerations of ethical and social implications are doomed to remain dry runs in hypothetical scenarios.

\section{Conclusion: Navigating the Future of AI within the EU AI Act Framework through HCI}

As we stand on the brink of a new era in Artificial Intelligence (AI) and Human-Computer Interaction (HCI), it is imperative to recognize the intertwined future of these disciplines under the legislative umbrella of the EU AI Act. The present paper explored the multifaceted implications and preconditions of the EU AI Act, emphasizing the pivotal role of HCI in fostering systems that are not only technologically advanced but also ethically aligned and human-centric.

The flourishing  landscape of AI, characterized by rapid advances and integration into everyday life, calls for re-evaluating the role of the field of HCI. That is, HCI must evolve from its traditional boundaries to address the complexities of how AI systems are intertwined with humans, with technology becoming increasingly autonomous in its information processing and action regulation yet still intimately linked to human activities. Tackling this (co-)evolution of a new human-technology relationship necessitates a collaborative, interdisciplinary approach that merges technical advances on the frontier of XAI with insights from a deep understanding of the psychology of Human-AI interaction closely integrated with broad perspective from social sciences, ethics, and law.

In light of the EU AI Act, this consequently underscores the necessity for HCI to be at the forefront of designing AI systems that prioritize transparency, accountability, and inclusivity. That is, the Act's regulatory framework provides a unique opportunity for HCI researchers and practitioners to lead the development of standards and methodologies that ensure AI systems are comprehensible and beneficial to all segments of society.

The future challenges and opportunities for HCI within this framework are manifold. They include crafting interfaces that enable meaningful human oversight, developing evaluation methodologies that encompass ethical considerations, and ensuring that AI systems enhance rather than diminish human capabilities. Furthermore, the role of HCI in education and public engagement is critical for demystifying AI technologies and fostering a society that is informed, prepared, and optimistic about the AI-driven future.

We conclude with a call to action for the HCI community to proactively engage with the challenges posed by the AI revolution. By embracing the principles of human-centered design, interdisciplinary collaboration, and ethical responsibility, HCI can lead the way in ensuring that AI technologies are developed and deployed in a manner that truly benefits humanity. In doing so, HCI will not only respond to the immediate demands of the EU AI Act but also shape the long-term trajectory of AI and HCI for a future where technology serves to amplify human potential and societal well-being.

\bibliographystyle{apalike}
\bibliography{localreferences,references}

\section*{Acknowledgement}
Part of this research was funded by the German Federal Ministry for Family Affairs, Senior Citizens, Women and Youth (BMFSFJ) under the funding program “Richtlinie zur Förderung von Künstlicher Intelligenz für das Gemeinwohl“, project Wegweiser.UX-für-KI: Online-Kompetenzaufbau "UX für gemeinwohlorientierte Kl" (FKZ 3923406K02).

\end{document}